\documentclass{camera}
\usepackage[final]{graphicx}  
\usepackage{amsmath}
\begin{document}

 
\begin{figure}
\begin{center}
    \includegraphics[width=0.21\textwidth]{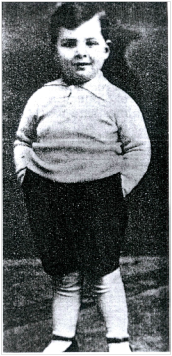}
    \includegraphics[width=0.305\textwidth]{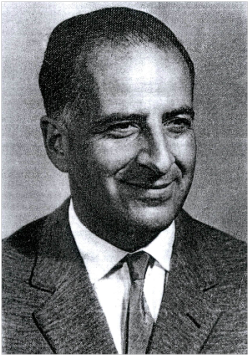}
\end{center}
\end{figure}

\title{On the history of the PMNS Matrix\\
...with today's perspective}
\author{J. BERNABEU}

%
\organization{Department of Theoretical Physics, University of Valencia,\\
and IFIC, Univ. Valencia-CSIC, E-46100 Burjassot, Valencia}

\maketitle

\begin{abstract}
The conceptual basis for understanding the interplay of neutrino mass and mixing for neutrino oscillations were paradoxically discussed in a period when the prevailing view was that of massless neutrinos. The name of Bruno Pontecorvo is associated to most of the components for this beautiful quantum phenomenon: muon-electron universality, different neutrino flavours, mismatch between weak interaction and mass eigenstates, neutrino oscillation phenomenology, including flavour and Majorana transitions.
\end{abstract}

%

\section{Personal recollections}

I met personally Bruno Pontecorvo in the summer of 1990, when he visited
CERN at the time of the collapse of the former Soviet Union. He was deeply
concerned and wishful on the future of Russia and I remember our
conversations following the news by radio.

\begin{figure}
  \centering
    \includegraphics[width=0.50\textwidth]{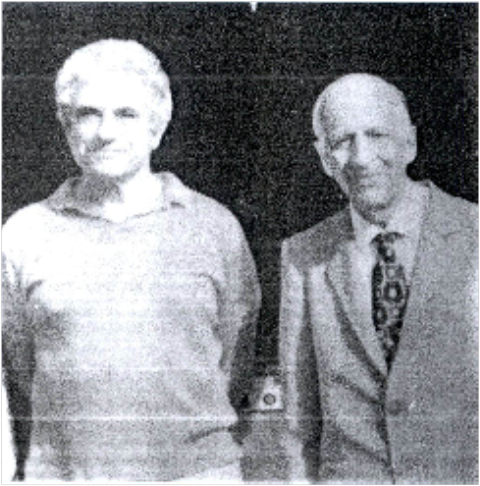}
  \caption{Bruno with Jack Steinberger at CERN}
  \label{fig:1}
\end{figure}

Beyond political events, his stay at CERN stimulated joint meetings among
physicists interested in neutrino physics. In fig.1 we see a picture of
Bruno with Jack Steinberger in these days. On the other hand, I was
delighted in convincing Bruno to participate in TAUP'91 workshop in Toledo
and NEUTRINO'92 conference in Granada. 

According to notes by Franco Buccella in Pontecorvo'
s book \cite{Referencia1}, when Bruno joined the TAUP meeting in the Lecture Hall of
Toledo, I was lecturing on "Neutrino Properties" \cite{Referencia2} and interrupted my
presentation with the greeting "Bruno, Welcome to Spain". After a moment of
general complacency, the session continued. The participation of Bruno
Pontecorvo in Spanish events of physics had a new glorious point with the
NEUTRINO'92 conference \cite{Referencia3} in Granada and the Universal Exhibition in
Sevilla. In fig. 2  we see Bruno in the social dinner of the
conference and enjoying the after-dinner performance of flamenco dancing.
Among other exhibitions in Sevilla, the Canada Pavilion,
was special for neutrino physicists with the presentation of the SNO
observational proposal for solar neutrino detection. The provocative
statement was: "John Bahcall is probably right. But his solar model will not
be needed for the interpretation of the solar neutrino problem". Bruno
Pontecorvo had envisaged  \cite{Referencia4} the solar neutrino problem by
predicting neutrino oscillations. The SNO experiment \cite{Referencia5} was able to
disentangle this particle physics solution from the astrophysical solution
through the comparison of neutrino fluxes at the detector as measured from
charged-current and neutral-current reactions on deuterium.

\begin{figure}
\begin{center}
  \includegraphics[width=0.45\textwidth]{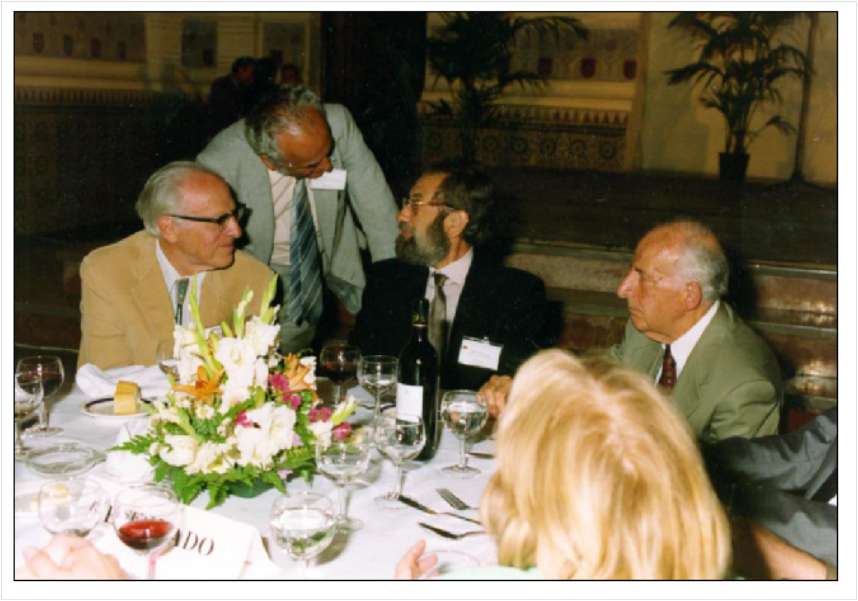}
    \includegraphics[width=0.45\textwidth]{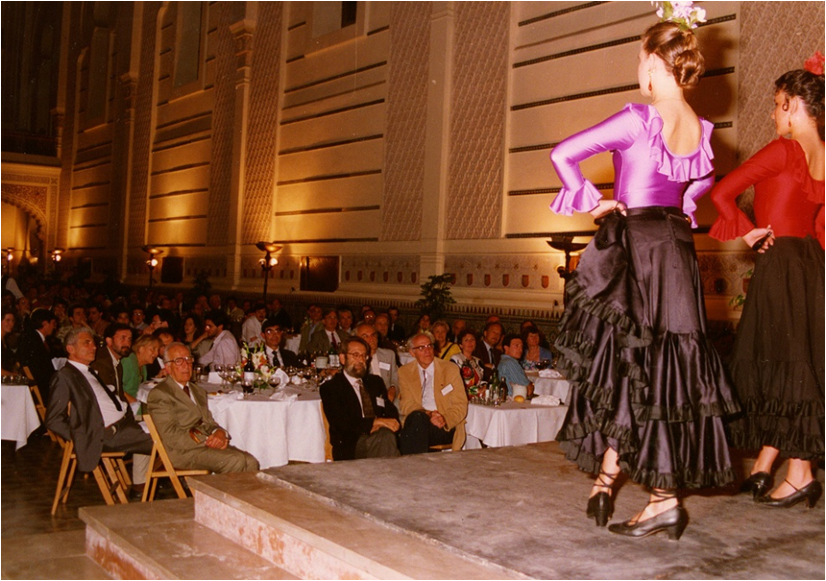}
\caption{Bruno in the social dinner and the after-dinner performance at NEUTRINO'92}
\end{center}
\end{figure}

CERN, as Meeting Point of physicists, was also instrumental in preparing a
long-term visit of Samoil Bilenky, from 1991 to 1994, to Valencia. This
period was very fruitful in scientific collaborations and generated a deep
friendship which is lasting until today. In fig. 3 we see a picture of
Samoil with Bruno in a moment of physics discussions.

\begin{figure}
\begin{center}
    \includegraphics[width=0.45\textwidth]{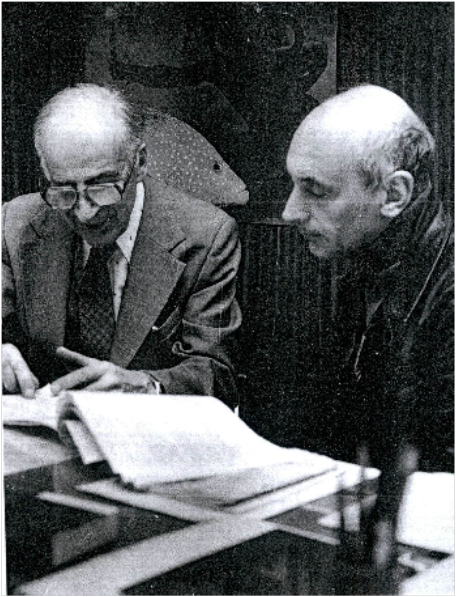}
\caption{Bruno with Samoil Bilenky}
\end{center}
\end{figure}

\section{The components of neutrino mixing}

The understanding of the beautiful properties associated to Neutrino Mixing
and Oscillations includes several "components" which we discuss in their
historical steps:

\begin{enumerate}
\item {The Lepton Family Problem}

\begin{itemize}
\item {$\mu$-e Universality} 

\item {Different $\nu _{e}-\nu _{\mu }$ Flavours}
\end{itemize}

\item {Neutrino Mass}

\begin{itemize}
\item {Mismatch between Weak Interaction-Mass Eigenstates} 

\item {Global L-charge?}
\end{itemize}

\item {Mixing and Oscillations}

\begin{itemize}
\item {Earliest ideas} 

\item {MNS mixing in the Nagoya model of baryon structure} 

\item {Oscillation Phenomenology}
\end{itemize}
\end{enumerate}

\section{The Lepton Family Problem}

\subsection{$\mu$-e Universality} 

A decade before the (V-A) theory of (charged
current) weak interactions, Bruno Pontecorvo discussed \cite{Referencia6} the
"universality" of weak interactions for processes of nuclear $\beta$-decay
together with those with muon and neutrino.

The process with the muon-neutrino pair is muon-capture

\begin{equation}\label{Ecuacion1}
\mu^{-}+(A,Z) \rightarrow  \nu + (A,Z-1)
\end{equation}

Following the indication given by the result of the experiment by Conversi,
Pancini and Piccioni, B. Pontecorvo compared the probability of this process
with the probability of the K-capture

\begin{equation}
e^{-} + (A,Z) \rightarrow  \nu + (A,Z-1)
\end{equation}

He came to the conclusion that the coupling constant of the interaction of
the muon-neutrino pair with nucleons is of the same order as the Fermi
coupling constant for $\beta$-decay and e-capture.

\begin{figure}
\begin{center}
\includegraphics[width=0.50\textwidth]{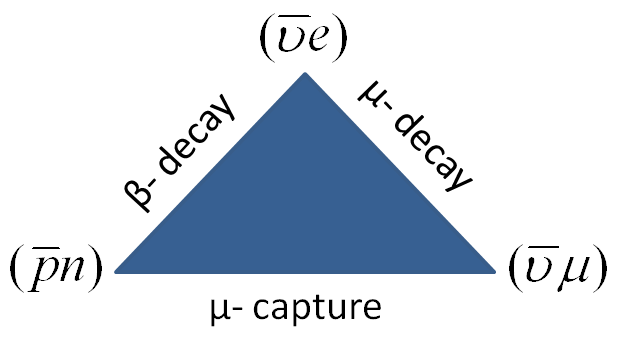} 
\caption{The Puppi triangle}
\end{center}
\end{figure}

The idea of $\mu$-e universality of the weak interaction was also followed
by G. Puppi \cite{Referencia7}. Puppi presented it in the form of a triangle, the "Puppi
triangle" of fig. 4, assuming that a universal weak interaction includes not
only the Hamiltonians of the $\beta$-decay and mu-capture but also the
Hamiltonian of the $\mu$-decay

\begin{equation}
\mu ^{+}\rightarrow e^{+}+\nu +\overline{\nu }
\end{equation}
Puppi suggested that the different parts of the weak interaction were the
sides of a triangle with vertices

\begin{equation}
\left( \bar{p}n\right) -\left( \bar{\nu }e\right) -\left(
\bar{\nu }\mu \right)
\end{equation}
and the Hamiltonian is given by a sum of products of different vertices, the
"currents".
A question was still open: Is the same $\nu$ in the two vertices of fig. 4?

\subsection{The Lepton Flavour Number}

The idea of different neutrinos $\nu _{e} - \nu _{\mu }$ appeared published in
a paper by B. Pontecorvo "Electron and Muon Neutrinos" \cite{Referencia8}. Even more
important, the concept of the Brookhaven experiment that discovered the muon
neutrino was due to B. Pontecorvo \cite{Referencia9} in 1959.

A direct proof of the existence of the second (muon) neutrino was obtained
by Lederman, Schwartz, Steinberger et al. in the first experiment with
accelerator neutrinos in 1962 \cite{Referencia10}. This discovery was a great event in
physics: the concept of Lepton Flavour Number thus appeared, with $L_{e}$
for $e^{-}$ and $\nu_e$ and $L_{\mu }$ for $\mu ^{-}$ and $\nu _{\mu }$.

The neutrino beam in the Brookhaven experiment was practically a pure $\nu_{\mu}$ beam from $\pi^+$ decay, with a small admixture of $\nu_{e}$ from
decays of muons and Kaons. The $\nu _{\mu }$ produces $\mu^{-}$ in the process

\begin{equation}
\nu _{\mu }+N\rightarrow \mu ^{-}+X
\end{equation}

If $\nu _{\mu }$ and $\nu _{e}$ were the same particle, neutrinos from  $\pi^+$  decay
would also produce $e^-$ in the process

\begin{equation}
\nu _{\mu }+N\rightarrow e ^{-}+X
\end{equation}

Due to the $\mu^{-}e$ universality of the weak
interaction, one would expect to observe in the detector practically equal
numbers of muons and electrons. In the Brookhaven experiment 29 muon events
were detected. The observed 6 electron candidates could be explained by the
background.

The measured cross section was in agreement with the V-A theory. Thus, it
was proved that $\nu _{\mu}$ and $\nu _{e}$ are different particles.

The total electron and muon lepton numbers $L_{e}$ and  $L_{\mu}$ are separately conserved by weak interactions 

\begin{equation}
\sum_{i}L_{e}^{\left( i\right) }=const; \,\,\,\,\,\,\,\,\, \sum_{i}L_{\mu}^{\left( i\right) }=const
\end{equation}

The flavour lepton numbers of the particles are given in Table 1.\\ 
The lepton numbers of antiparticles are opposite to the lepton numbers of the corresponding particles

\begin{table}[h]
\centering
\caption{}
\begin{tabular}{cccc}
\hline
Lepton number & $\nu_{e}$, $e^{-}$ & $\nu_{\mu}$, $\mu^{-}$ & hadrons, $\gamma$ \\
\hline
$L_{e}$ & 1 & 0 & 0 \\
$L_{\mu}$ & 0 & 1 & 0 \\
\hline
\end{tabular}
\end{table}

An earlier indication that $\nu_{e}$ and  $\nu_{\mu}$ are different particles was obtained from the data on the search for the decay 
$ \mu \rightarrow  {e}  \gamma $.  If $\nu_{\mu}$ and $\nu_{e}$ are identical particles, this decay is allowed. The probability of the decay in the theory with the W-boson was calculated  by G. Feinberg\cite{Referencia11}, finding that the branching ratio R of the radiative decay to the ordinary muon decay would be $R \sim 10^{-4}$. At the time of the Brookhaven experiment, an experimental upper bound $R<10^{-8}$ had been found.

\section{Neutrino mass}

The history of the neutrino mass problem is one of Up´s and Down´s.The phenomenon of Parity Violation in processes involving neutrinos led to the advent of the "Two-component neutrino theory". In terms of the chiral components of the neutrino field, left-handed $\nu_{L}(x)$ and right-handed $\nu_{R}(x)$, the Dirac equation is written

\begin{equation}
i\gamma ^{\mu }\partial _{\mu }\nu _{L}\left( x\right) -m_{\nu }\nu_{R}\left( x\right) =0
\end{equation}

If neutrinos are exactly massless the two chiral components are decoupled, so that the door is open to a definite chirality-helicity. The spectacular Goldhaber experiment \cite{Referencia12} determined the neutrino helicity to be left-handed. It was obtained, using conservation of angular momentum only, from the measurement of the circular polarization of the photon emitted in the nuclear transition of the final $^{132}$Sm in the electron capture reaction by $^{152}$Eu

\begin{eqnarray}
e^{-}+^{152}{E}u\rightarrow\nu +\underset{\downarrow }{^{152}{Sm}^*}  
\nonumber\\^{152}{Sm+\gamma} 
\end{eqnarray}

The spins of $^{152}$Eu and $^{152}$Sm are equal to zero and the spin of $^{152}$Sm$^{*}$ is equal to one. For K-capture, the circular polarization of the $\gamma$'s emitted in the direction of the $^{152}$Sm$^{*}$ recoil momentum is equal to the neutrino helicity. The result was compatible with 100$\%$  negative helicity of the neutrino emitted in electron capture. The direct measurement \cite{Referencia13} of the helicity for the muonic neutrino was performed later following the restrictions \cite{Referencia14} imposed for the recoil polarization in the muon capture process

\begin{equation}
\mu^{-}+^{12}C \rightarrow \nu +^{12}B
\end{equation}

However, the Universal V-A theory of weak interactions tells us that the left-handed chiral fields enter for all elementary fermions, not only for neutrinos. As a consequence, there is no rationale why neutrinos should be special and massless. Still, the difference is that the other elementary fermions have an electric charge (and gluonic colour for quarks) and parity conserving electromagnetic (and colour for quarks) interactions, so that the opposite helicity component has to exist and it is active. Neutrinos have neither electric charge nor colour charge. Do they have a Global Lepton Number distinguishing neutrinos and antineutrinos? Still in 2013, this is an open question.

Already in 1946, B. Pontecorvo made the proposal \cite{Referencia15}: For $\overline{\nu }$'s produced by $\beta$-decay in nuclear reactors, can they produce $e^{-}$'s? This problem was studied in an experiment which was performed in 1956 \cite{Referencia16} by Davis at the Savannah River reactor. This was in fact the first application of Pontecorvo's radiochemical method. Radioactive $^{37}Ar$ atoms produced in the process

\begin{equation}
\bar{\nu}+^{37}Cl\rightarrow e^{-}+^{37}Ar
\end{equation}
were searched for in the experiment. No $^{37}Ar$ atoms were found. The experimental upper bound for the cross section was a factor five smaller than the corresponding value for neutrinos. Thus, it was established that antineutrinos from a reactor can produce positrons (the Reines-Cowan experiment) but they can not produce electrons (the Davis experiment). We can assign an additive global lepton charge to these interacting neutrinos by weak interactions.
Is it possible for neutrinos to acquire a Majorana mass \cite{Referencia17}? Forbidden for the other elementary fermions due to exact electric charge conservation, for neutrinos it is a priori allowed iff the mass terms violate global lepton charge by two units. In this case, the states of neutrinos with definite Majorana mass would be a linear superposition of weak interacting neutrinos with opposite lepton charge.

Global Lepton Number would then be not defined for neutrinos with definite Majorama mass. Even more: one can have massive neutrinos with the active (left-handed) chiral component only and the sterile (right-handed) component is not needed. Contrary to Dirac fermions, Majorana fermions have two degrees of freedom, the neutrino of left-handed chirality and its conjugate. The states of definite mass and helicity, which are compatible observables, are the left-handed with a relative m/E component of the conjugate and its orthogonal.

\section{Neutrino mixing and oscillations}

\subsection{Early ideas}

Already in 1957, B. Pontecorvo writes \cite{Referencia18}: "If the theory of two component neutrino was not valid, and if the conservation law for "neutrino charge" took not place, neutrino $\longleftrightarrow$ antineutrino transitions would be possible". In this statement one finds the two essential ingredients for oscillations: neutrino mass and mixing. In these early ideas, Pontecorvo discussed oscillations in analogy with Gell-Mann \& Pais theory of $K^o  - \bar{K}^o$ mixing and oscillations.

Instead of having the active neutrinos only $\nu_{L}$ and $\left(\bar{\nu}\right)_{R}$, Pontecorvo assumed additional neutrinos $\left(\bar{\nu}\right)_{L}$ and $\nu_{R}$ with the name of "sterile" neutrinos. In connection with the Davis experiment \cite{Referencia16}, he considered the active-sterile mixing  $\left(\bar{\nu}\right)_{R}$ $\longleftrightarrow$ $\nu_{R}$, with two massive Majorana states 

\begin{equation}
\nu _{1}=\frac{1}{\sqrt{2}}\left[ \left( \bar{\nu }\right) _{R}+\nu_{R}\right],\,\, \nu _{2}=\frac{1}{\sqrt{2}}\left[ \left( \bar{\nu }\right) _{R}-\nu
_{R}\right] 
\end{equation} with a mass difference $\Delta$m. Pontecorvo obtained the neutrino oscillation results \cite{Referencia19} 

\begin{equation}
Appearence\,\, P\left[ \left( \bar{\nu }\right) _{R}\xrightarrow{L}\nu_{R}\right] =\frac{1}{2}\left( 1-\cos \frac{\Delta m^{2}L}{2E}\right) \,\,\,   Davis
\end{equation}
\begin{equation}
Survival \,\, P\left[ \left( \bar{\nu }\right) _{R}\xrightarrow{L}\left( \overline{\nu }\right) _{R}\right] = 1-P\left[ \left( \bar{\nu }\right) _{R}\xrightarrow{L}{\nu }_{R}\right]\,\, Reines-Cowan
\end{equation}

The result of eq. (13) was of relevance for the Davis experiment, that of eq. (14) for the Reines-Cowan experiment. Pontecorvo writes \cite{Referencia19}: "It would be extremely interesting to perform the Reines-Cowan experiment at different distances L from the reactor". Such experiments were performed in the last decades and it was only in 2003 that KamLAND \cite{Referencia20} observed for the first time the oscillation effect with reactor active antineutrinos.

\subsection{Neutrino Mixing for Baryon Model}

In 1962, the MNS paper "Remarks on the Unified Model of Elementary Particles" appeared \cite{Referencia21}. What was the "unified" Model? It refers to the Nagoya model of Baryons as bound states of neutrinos and "a new sort of matter" vector boson. In order to explain the smallness of the leptonic decay rate of hyperons, the "true neutrinos" in these baryons would be

\begin{equation}
\nu _{1}=\cos \delta\,\,\, \nu _{e}-\sin \delta \,\,\,\nu _{\mu },\,\, \,\,\,\, 
\nu _{2}=\sin \delta \,\,\,\nu _{e}+\cos \delta \,\,\,\nu _{\mu }
\end{equation}

As a consequence, $\delta$ should be identified with the Cabibbo angle.

The MNS neutrino mixing was not associated to the quantum phenomenon of neutrino oscillations, with interference between the neutrino mass eigenstates. The state $\nu_{2}$, on the contrary, would have additional interaction with a field of heavy particles X. In MNS words, "Weak neutrinos are not stable due to the occurrence of virtual transitions $\nu_e \leftrightarrow \nu_{\mu}$ caused by this additional interaction with $\nu_{2}$".

\subsection{Neutrino Oscillation Phenomenology}

After the discovery of the muonic neutrino $\nu_{\mu}$, in 1967 Pontecorvo discussed \cite{Referencia4} the phenomenology of neutrino oscillations in modern views, including the flavour transitions $\nu_e \leftrightarrow \nu_{\mu}$ and the Majorana transitions $\nu_e \leftrightarrow \left(\bar\nu_e\right)_ L$ and $\nu_{\mu} \leftrightarrow \left( \bar\nu_{\mu}\right)_ L$. Among other subjects, he applied this study to solar neutrino oscillations.

At that time R. Davies started his famous experiment on the detection of solar neutrinos in which the radiochemical method of neutrino detection, proposed by B. Pontecorvo in 1946, was used. Solar neutrinos were detected in this experiment via the observation of the reaction

\begin{equation}
\nu_e +^{37}Cl\rightarrow e^{-}+^{37}Ar
\end{equation}

The results \cite{Referencia22} created "the solar neutrino problem". In a sense, Pontecorvo had envisaged the existence of this problem.

In the paper by V. Gribov and B. Pontecorvo \cite{Referencia23} in 1969, one reads:"If Global Lepton Number is violated, neutrinos would have a mass of Majorana type". The scheme of two neutrino mixing proposed by them was the minimal one. In this scheme, the only possible oscillations are  $\nu_e \leftrightarrow \nu_{\mu}$, there are no sterile neutrinos and the four states of flavour neutrinos and antineutrinos form the states of two massive Majorana neutrinos with helicities $\pm$1. The effect of these vacuum oscillations on the flux of solar neutrinos on the earth was discussed.

S. Bilenky and B. Pontecorvo [24] introduced the neutrino mixing between the two families on the basis of the lepton-quark analogy. They discussed possible neutrino oscillations in reactor and accelerator neutrino experiments. For more than two neutrinos, N. Cabibbo studied \cite{Referencia25} the requirements for invariance$\backslash$non-invariance of CP and T symmetries.

In 1998, in the Super-Kamiokande atmospheric neutrino experiment \cite{Referencia26}, a significant up-down asymmetry of the high-energy muon events was observed. In this way it was proved that the number of observed muon neutrinos depends on the distance which neutrinos passed from a production point in the earth atmosphere to the detector. The Super-Kamiokande atmospheric neutrino result was the first model independent evidence of neutrino oscillations. The Golden Years of neutrino oscillation physics started with this fundamental discovery.

\section{Conclusion}

The Discovery of neutrino oscillations in 1998, implying neutrino mass differences and neutrino mixing, was a great event in Science. In the last 15 years, the progress in the determination of these neutrino properties has been impressive from atmospheric, solar, reactor and accelerator experiments. A general recent review on neutrino physics covering all aspects of the field can be seen in a special issue of Adv. High En. Phys.\cite{Referencia27}. With today's perspective, we condense the information in the Unitary Mixing Matrix for three active neutrinos $\nu _{e}\Leftrightarrow \nu _{\mu };\,\,\,\,\nu _{e}\Leftrightarrow \bar{\nu }_{e};\,\,\,\,\nu _{\mu }\Leftrightarrow \bar{\nu }_{\mu }$

\begin{equation}
U\!\!=\!\!\! \left[ \!\!\!
\begin{array}{ccc}
1 & 0 & 0 \\ 
0 & c_{23} & s_{23} \\ 
0 & -s_{23} & c_{23}
\end{array}
\!\!\!\right]\!\!\!\! \left[\!\!\!
\begin{array}{ccc}
c_{13} & 0 & s_{13}e^{-i\delta } \\ 
 0 & 1 & 0 \\ 
-s_{13}e^{i\delta } & 0 & c_{13}
\end{array}
\right] \!\!\!\! \left[\!\!\! \begin{array}{ccc}
c_{12} & s_{12} & 0 \\ 
-s_{12} & c_{12} & 0 \\ 
0 & 0 & 1
\end{array}
\right] \!\!\!\!\left[ \!\!\!
\begin{array}{ccc}
e^{i\alpha _{1}/2} & 0 & 0 \\ 
0 & e^{i\alpha _{2}/2} & 0 \\ 
0 & 0 & 1
\end{array}
\!\!\!\right] 
\end{equation}
\\
In this form the mixing matrix is valid for both Dirac and Majorana neutrinos. For flavour oscillations, the last diagonal matrix of phases is unobservable. The "natural" parameterization of these phases for Majorana neutrino-antineutrino transitions is with $\frac{\alpha }{2}$ in the vertex, the physical reason being that $\alpha$ represents the relative CP-phase \cite{Referencia28} between two Majorana neutrinos. CP conservation corresponds to $\alpha$ = 0, $\pi$ ,i.e., a Majorana neutrino with relative CP-eigenvalue equal to +1, -1.

Historically \cite{Referencia29}, it is spectacular that the concepts involved in the problems of neutrino mass, mixing and oscillations were discussed, and understood, in a period when the prevailing view was that of massless neutinos. The essential steps discussed in this paper were: e-$\mu$ universality, different neutrino families, interplay of mass and
mixing for neutrino oscillations, neutrino flavour mixing for baryon structure, neutrino mixing for oscillation phenomenology including flavour and Majorana transitions. In all these conceptual basis, but one, the name of Bruno Pontecorvo appears as the prominent discoverer. My conclusion is that it is fair to call the U matrix 
\\

 {\LARGE {The PMNS Matrix}}

\section*{Acknowledgments}
The Pontecorvo100-Symposium was a memorable event in honour of Bruno Pontecorvo. I would like to thank the organizers for the privilege of participating in the Symposium with my contribution. Suggestions and advice by Samoil Bilenky are specially acknowledged. This research has been supported by MINECO with Grant FPA-2011-23596 and Generalitat Valenciana with Grant PROMETEOII-2013-017.



%

\end{document}